\RequirePackage{ifpdf}
\ifpdf
\documentclass[a4paper,12pt,twoside]{article}
\else
\documentclass[a4paper,12pt,twoside,dvips]{article}
\fi

\usepackage[scale=0.85,includeheadfoot]{geometry}
\usepackage[medium]{titlesec}

\usepackage[utf8x]{inputenc}
\usepackage[english]{babel}

\usepackage[T1]{fontenc}

\usepackage{amsmath,amsfonts,amsthm, mathtools}
\mathtoolsset{showonlyrefs,showmanualtags}
\usepackage{subeqnarray}
\usepackage{latexsym}

\usepackage{mathptmx}

\usepackage{parskip}
\usepackage{fancyhdr}
\usepackage{graphicx}
\usepackage{hyperref}
\usepackage{subeqnarray}

\graphicspath{{pics/}}

\renewcommand{\Box}{\#}

\newcommand{\mathset}[1]{\mathbb{#1}}

\newcommand{\R}{\mathset R}

\newcommand{\Id}{  1 \!\!\! \mathrm I}

\newcommand{\T}{{\mathbb T}}
\newcommand{\dd}{{\mathrm d}}
\newcommand{\trace}{\mathrm{tr}}

\newcommand{\OPS}{\; \mathrm{OPS}}

\theoremstyle{plain}

\theoremstyle{definition}

\theoremstyle{remark}

\title{Efficient Higher Order Derivatives of Objective Functions Composed of Matrix Operations}

\author{Sebastian~F. Walter}

\begin{document}
\maketitle

\tableofcontents

\section{Introduction}
This paper is concerned with the efficient evaluation of higher-order derivatives of functions $f$ that are composed of matrix operations. I.e., we want to compute the $D$-th derivative tensor
\begin{equation}
\nabla^D f(X) \in \R^{N^D}\;,
\end{equation}
where $f:\R^{N} \rightarrow \R$ is given as an algorithm that consists of many matrix operations. We propose a method that is a combination of two well-known techniques from Algorithmic Differentiation (AD):
univariate Taylor propagation on scalars (UTPS) \cite{adbook,exactinterpolation} and first-order  forward and reverse on matrices~\cite{giles:08}. The combination leads to a technique that we would like to call univariate Taylor propagation on matrices (UTPM).
The method inherits many desirable properties: It is easy to implement, it is very efficient and it returns not only  $\nabla^D f$ but yields in the process also the derivatives $\nabla^d f$ for $d \leq D$. As performance test we compute the gradient $\nabla f(X)$ 
of $f(X) = \trace ( X^{-1})$ in the reverse mode of AD for $X \in \R^{n \times n}$. We observe a speedup of about $100$ compared to UTPS. Due to the nature of the method, the memory footprint is also small and therefore can be used to differentiate functions that are not accessible by standard methods due to limited physical memory.

The following sections are structured as follows: In Sect. \ref{sec:ad_introduction} we give a brief explanation of the key ideas of AD. In Sect. \ref{sec:UTPS} we give a summary of UTPS which is then used in Sect. \ref{sec:ad_modes} where the forward and reverse mode of AD are explained. In Sect. \ref{sec:combination} we show how the forward and reverse mode can be combined to compute higher order derivatives. Sect. \ref{sec:intro_matrix_ad} serves as motivation for UTPM. In Sect. \ref{sec:utpm} the central idea of UTPM is introduced. Section \ref{sec:utpm_reverse} shows how this idea is applied to the reverse mode of AD followed by Section \ref{sec:utpm_higher} where the combination of forward and reverse mode on matrices is explained.
In Sect. \ref{sec:complexity} we briefly discuss the complexity of UTPM compared to UTPS and in Sect. \ref{sec:performance} we show how our proposed method performs compared to existing state-of-the-art methods in practice.

\section{Computation and Algorithmic Differentiation}
\label{sec:ad_introduction}
\begin{figure}[!ht]
\centering
\begin{minipage}{0.6\textwidth}
\includegraphics[angle=0, width=\textwidth]{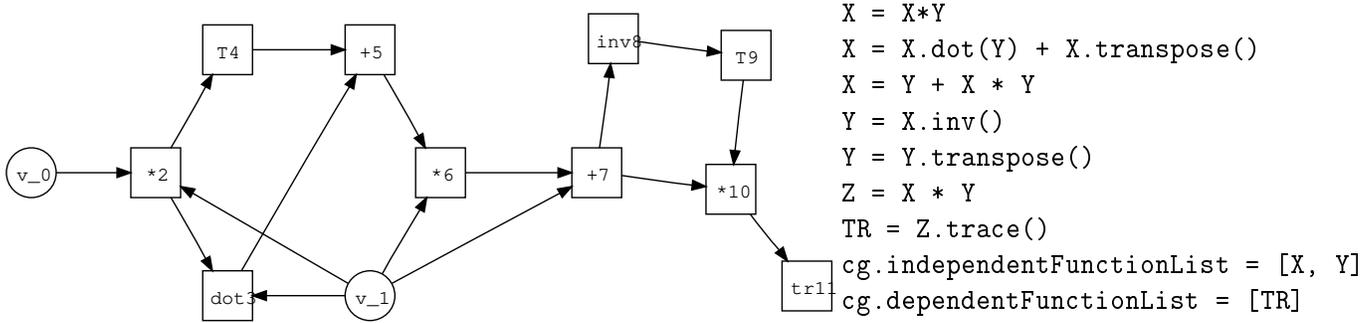}
\end{minipage}
\begin{minipage}{0.375\textwidth}
\vspace*{-0.0cm}
\small
\begin{verbatim}
X = X*Y
X = X.dot(Y) + X.transpose()
X = Y + X * Y
Y = X.inv()
Y = Y.transpose()
Z = X * Y
TR = Z.trace()
cg.independentFunctionList = [X, Y]
cg.dependentFunctionList = [TR]
\end{verbatim}
\end{minipage}
\vspace*{0cm}
\caption{\label{fig:example_cg}
The computational graph on the left side is defined by the computer program on the right side. The variables \texttt{X} and \texttt{Y} are matrices. The squares represent function nodes. The numbers represent the occurrence in the sequence of successive operations. Independent variables are represented as circles.}
\end{figure}

A \emph{program} is a sequence of instructions that a computer can interpret step by step.
Generally, functions of practical interest in science and engineering can be evaluated as a program.
Mathematically speaking, such functions are composite functions of elementary functions.
The definition of \emph{elementary} is not strict. In fact, only the four operators $+,-,\times,/$ are really elementary: they are required to define the field of real numbers $\R$. The theory of \emph{Algorithmic Differentiation} (AD) is the application of the chainrule to the sequence of elementary functions. In the context of AD, we mean by elementary functions such functions that have ``nice'' analytical properties, i.e. can be differentiated analytically.
 In formulas, we want to evaluate functions $F: \R^N \rightarrow \R^M$ that are built of elementary functions $\phi$:
\begin{eqnarray}
F : x &\mapsto& y = F(x) \;,
\end{eqnarray}
where $x \equiv (x_1,\dots, x_N),\; y \equiv (y_1,\dots, y_M)$. If $M=1$ we use $f$ instead of $F$.
For example $f(x_1,x_2) = x_1*x_2 + x_1^2$ can be written as
\begin{equation*}
f(x_1,x_2) = \phi_3( \phi_1(x_1,x_2), \phi_2(x_1)) = \phi_3(v_1,v_2).
\end{equation*}
We use the notation $v_l$ for the result of $\phi_l$ and $v_{j \prec l}$ for all arguments of $\phi_l$. To be consistent, the independent input arguments $x_n$ are also written as $v_{n-N} = x_n$. To sum it up, the following three equations describe the function evaluation:
\begin{subeqnarray}
v_{n-N} & = & x_n \quad\quad\quad n=1,\dots,N\\
v_{l} & = & \phi_l(v_{j \prec l}) \quad l =1,\dots, L\\
y_{M-m} & = & v_{L-m} \quad\quad m = M-1, \dots, 0 \;,
\end{subeqnarray}
where $L$ is the number of calls to elementary functions $\phi_l$ during the computation of $F$ ($L=3$ in the above example). Running indices ($n$,$m$, $l$) use the same letter as the boundary values ($N$,$M$,$L$) to make the notation easier to read. The sequence can also be represented by a computational graph, as depicted in Fig. \ref{fig:example_cg}.

To differentiate such a program given as sequence of elementary functions $\phi_l$ the \emph{chain rule} is applied to each elementary function $\phi_l$:
\begin{eqnarray}
\dd \phi_l( v_{j \prec l}) &=&  \sum_{j \prec l} \frac{\partial \phi_l}{\partial v_j} \dd \phi_j \;.
\end{eqnarray}
That means that the evaluation of the derivative of $F$ breaks down to differentiating the elementary functions $\phi_l$. In contrast to symbolic differentiation, rather than the symbolic expression, the numerical value is propagated. In the following sections we will concentrate on one elementary function $f \equiv \phi_l$ and keep in mind that we have then treated arbitrary functions that are composed of such elementary functions.

\section{Univariate Taylor Propagation on Scalars}
\label{sec:UTPS}
In this section it is explained how higher order derivatives of functions $F: \R^N \rightarrow \R^M$ can be computed by means of \emph{Univariate Taylor Propagation on Scalars} (UTPS). This theory has been successfully implemented in software by use of operator overloading, for example ADOL-C \cite{adolc:96} or CppAD \cite{cppad}.
The key observation is that the propagation of a univariate truncated Taylor polynomial $x_0 + t \in \mathbb T_D$ of degree $D$ through a function $f: \R \rightarrow \R$ yields the derivatives $\dd^d f$, $ 0 \leq d \leq D$:
\begin{eqnarray}
f(x_0 + t) &=& \sum_{d=0}^D \frac{1}{d!} \dd^d f(x_0) t^d + \mathcal O(t^{D+1}) \;. \label{eqn:der_from_utps}
\end{eqnarray}
In the application of the chain rule to the elementary functions $\phi_l$, the Taylor coefficients $v_d^l$ are filled with non-zero entries. In general, UTPS is given by
\begin{eqnarray}
 \sum_{d=0}^D y_d t^d  = f( \sum_{d=0}^D x_d t^d) &=& \sum_{d=0}^D \frac{1}{d!} \frac{\dd^d}{\dd t^d} f \left. ( \sum_{c=0}^D x_c t^c)  \right|_{t=0} t^d + \mathcal O(t^{D+1})  \;.
\end{eqnarray}
The explicit formulas of $y_d$ for $d = 0, \dots, D$ have to be calculated analytically. For some simple functions explicit expressions can be obtained: See Table \ref{tab:utps_binary} for some examples. To ease the notation we sometimes use $[ \cdot ]$ when we mean a univariate Taylor polynomial. E.g. $[x] := \sum_{d=0}^D x_d t^d$.
\begin{table}[!h]
\centering
\begin{tabular}{| c | l |}
\hline
$\phi(u,v)$  & $d = 0,\dots,D$\\
\hline
$u + cv$ & $\phi_d = u_d + c v_d$ \\
$u \times v $ & $\phi_d = \sum_{j=0}^d u_j  v_{d-j}$ \\
$u / v $ & $\phi_d = \frac{1}{v_0} \left[ u_d - \sum_{j=0}^{d-1} \phi_j v_{d-j} \right]$\\
\hline
\end{tabular}
\caption{\label{tab:utps_binary}
UTPS of the binary functions $\phi \in \{ +,\times, / \}$. This table summarizes how the Taylor coefficients $\phi_d$ in $\sum_{d=0}^D \phi_d t^d = \phi( \sum_{d=0}^D u_d t^d, \sum_{d=0}^D v_d t^d)$ are computed.
}
\end{table}

\section{The Forward and Reverse Mode on Scalars}
\label{sec:ad_modes}
To compute first order derivatives, it is favorable to use the \emph{reverse mode} of AD when $M<N$. However, this rule is only valid when a algorithmic complexity model is used that discards memory movements. In practice, memory movements are not only a minor correction to the actual runtime on a computer, but in fact a major contributor. The rule of thumb is therefore: if $M < 5 N$ then the reverse mode is most likely favorable.

The \emph{forward mode} propagates directional derivatives. I.e. it applies the chain rule starting at $\phi_0$. This can easily be done with UTPS as explained in the previous section.
The \emph{reverse mode} computes derivative vectors by applying the chain rule starting at $\phi_L$, i.e. compute
\begin{eqnarray}
\bar F^T \dd F &=& \bar F^T \dd F(x) = \bar F^T \sum_{n=1}^N \frac{\partial F}{\partial x_n} \dd x_n = \sum_{n=1}^N \bar x_n^T \dd x_n \;.
\end{eqnarray}
The recursion continues by applying the chainrule to $ \dd x_n =  \dd x_n(z)$. The recursion is stopped if $x_n$ is an independent variable.

\paragraph{Example} We want to compute the function $ f( g(x), y) = g(x) y = x^2 y $.
In the forward mode we compute the directional derivative $\frac{\partial f}{\partial x}(x,y) = [1,0] \nabla f(x,y)$ :
\begin{subeqnarray*}
\lbrack x] &=& [x,1] \;; \lbrack y] = [y,0] \\
g([x]) &=& [x]^2 = [x,1][x,1] = [x^2,2x]\\
f([g],[y]) &=& [g][y] = [x^2,2x][y,0] = [x^2 y , 2xy] \,
\end{subeqnarray*}
where $f_1 = 2xy$ is the wanted directional derivative.

In the reverse mode we compute:
\begin{subeqnarray*}
\dd f( g, y) &=& \left. \frac{\partial f}{\partial z} (z, y) \right|_{z = g(x)} \dd g + \frac{\partial f}{\partial y} \dd y \\
&=& \underbrace{y}_{=: \bar g} \dd g + \underbrace{g}_{\bar y} \dd y \\
&=& \underbrace{ \bar g 2 x}_{=: \bar x} \dd x + \bar y \dd y \;,
\end{subeqnarray*}
where the gradient of $f(x,y)$ and can be read from $\bar x$ and $\bar y$:
\begin{eqnarray*}
 \nabla_{(x,y)} f(x,y) = ( \bar x, \bar y )^T = ( 2 y x, x^2)^T \;.
\end{eqnarray*}

\section{Combining Forward and Reverse Mode}
\label{sec:combination}
To compute higher-order derivatives efficiently, one can combine forward and reverse mode.
The important observation is that one can differentiate functions $F : \T_D^N \rightarrow \T_D^M$ that propagate univariate Taylor polynomials in the forward mode once more in the reverse mode. In consequence one obtain obtains derivatives of degree $D+1$.
The combination relies on the interchangeability of the differential operators $\dd$ and $\frac{\dd}{\dd t}$:
\begin{subeqnarray}
\dd F([x]) &=&  \sum_{d=0}^D \frac{1}{d!} \dd \frac{\dd^d}{\dd t^d} F \left. ( \sum_{c=0}^D x_c t^c)  \right|_{t=0} t^d  = \sum_{d=0}^D \frac{1}{d!} \frac{\dd^d}{\dd t^d} \underbrace{\dd F}_{=:G} \left. ( [x])  \right|_{t=0} t^d  \;,
\end{subeqnarray}
i.e. to compute one higher order of derivatives with the reverse mode one can symbolically differentiate $F$ to obtain $G = \dd F$ and then use UTPS on $G$. That we obtain one higher order of derivatives can be seen from Eqn. (\ref{eqn:der_from_utps}).

\paragraph{Example}
The goal is to compute the Hessian-vector product $H\cdot v$ at $x = (2,3,7)^T$ with $v = (1,0,0)^T$. The Hessian is defined by the function
\begin{eqnarray*}
f: \mathbb R^3 &\longrightarrow& \mathbb R\\
x &\mapsto& y = f(x) = x_1  x_2  x_3 \;
\end{eqnarray*}
and reads
\begin{equation*}
H = \left( \begin{matrix}
            0 & x_3 & x_2 \\
	    x_3 & 0 & x_1\\
	    x_2 & x_1 & 0\\
           \end{matrix}
	\right) \;.
\end{equation*}
I.e., we want to compute the first column $(0, x_3, x_2) = (0, 7, 3)$ of the Hessian.
\begin{center}
\begin{tabular}{|l c l c l c l|}
\hline
 $[v_{-2}]$ &=& $[ x_1]$ &=& $ [2,1]$&&\\
 $[v_{-1} ]$ &=& $[ x_2]$ &=& $ [3,0] $&&\\
 $[v_{0} ]$ &=& $ [ x_3]$ &=& $ [7,0]$&&\\
\hline
 $[v_{1} ]$ &=& $ [ v_{-2}] [v_{-1}]$ &=&$ [2,1][3,0]$  &=&$ [6,3]$ \\
 $[v_{2} ]$ &=& $ [ v_{1}] [v_{0}]$ &=&$[6,3][7,0]$ &=&  $ [42,21]$\\
\hline
$[ \bar v_{2}]$  &=& $ [ \bar y]$ &=&$[1,0]$&&\\
\hline
$[ \bar v_{1}]$  &=& $ [ \bar v_{2}] [v_{0}]$ &=&$[1,0][7,0]$&$=$&$[7,0]$\\
$[ \bar v_{0}]$  &=& $ [ \bar v_{2}] [v_{1}]$ &=&$[1,0][6,3]$&$=$&$[6,3]$\\
$[ \bar v_{-1}]$  &=& $ [ \bar v_{1}] [v_{-2}]$ &=&$[7,0][2,1]$&$=$&$[14,7]$\\
$[ \bar v_{-2}]$  &=& $ [ \bar v_{1}] [v_{-1}]$ &=&$[7,0][3,0]$&$=$&$[21,0]$\\
\hline
$[ \bar x]$  &=& $ [ \bar v_{-2}]$ &=&$[21,0]$&&\\
$[ \bar y]$  &=& $ [ \bar v_{-1}]$ &=&$[14,7]$&&\\
$[ \bar z]$  &=& $ [ \bar v_{0}]$ &=&$[6,3]$&&\\
\hline
\end{tabular}
\end{center}
The brackets $[x_1]$ denote truncated Taylor series. The purpose of the first three lines is solely to make the notation consistent. The next two lines are the FDE where the multiplication between truncated Taylor series as explained in Table \ref{tab:utps_binary} has been used. Then the first adjoint variable $[\bar y]$ is defined. From there, the adjoints are computed in reverse order. Finally, in the last three lines the adjoint variables are renamed. The first Taylor coefficient of $x,y,z$ are the first column of $H$, i.e. $(H_{11},H_{21}, H_{31}) = (\bar x_1, \bar y_1, \bar z_1)$

We obtain 
\begin{subeqnarray*}
\dd [f] &=&  \dd \sin ( [y]) \\
&=& \cos ([y]) \dd [y] \\
&=& \underbrace{[ \cos(y_0), - \sin(y_0) y_1 ]}_{=: [\bar y ]} \dd [y] \\
&=& [ \bar y ] \dd \exp( [x]) \\
&=& [\bar y ] \exp( [x]) \dd [x] \\
&=& [\bar y ] [\exp( x_0), \exp(x_0) x_1] \dd [x] \\
&=& [\cos(y_0) \exp(x_0), \cos(y_0) \exp(x_0) x_1  - \sin(y_0) y_1 \exp(x_0)] \dd [x] \\
&=& [ \bar x] \dd [x] \;,
\end{subeqnarray*}
where we find that $[ \bar x ] = $

\section{Algorithmic Differentiation on Matrices} \label{sec:intro_matrix_ad}
The theory of matrix differential calculus is well-known in the statistics and econometrics community and there are a number of textbooks and papers available, e.g. \cite{neudecker:88, minka:00} and references therein. Our work is based on the tutorial paper \emph{Collected Matrix Derivative Results for Forward and Reverse Mode Algorithmic Differentiation} by M. Giles \cite{giles:08}.
The need for higher order derivatives of matrix operations arises for example in optimal experimental design (OED) problems. The OED objective function $\Phi$ is a function that depends on the covariance matrix $C \in \R^{N_p \times N_p}$ of the parameters $p \in \R^{N_p}$. The covariance matrix $C$ is itself a complicated expression in $J=(J_1,J_2)$, where $J_1 \in \R^{N_M \times N_p}$ is the sensitivity of the measurement model functions and $J_2 \in \R^{N_C \times N_p} $ the sensitivity of the  constraint functions w.r.t. the parameters $p$.
In particular, the following NLP has to be solved w.r.t. the control variables $q$:
\begin{eqnarray*}
q_* &=& {\rm argmin}_{q \in S \subset \R^{N_q}} \Phi(C(J(q)))\nonumber \;,\\ 
\mbox{where} \quad C & = & \left( \begin{array}{cc} I & 0 \end{array} \right) 
\left( \begin{array}{cc} J_1^T J_1 & J_2^T \\ 
J_2 & 0 \end{array} \right)^{-1}
\left( \begin{array}{cc} J_1^T J_1 & 0 \\ 0 & 0 \end{array} \right)
\left( \begin{array}{cc} J_1^T J_1 & J_2^T \\ 
J_2 & 0 \end{array} \right)^{-T}
\left( \begin{array}{c} I \\ 0 \end{array} \right).
\end{eqnarray*}
Typical NLP solvers need at least gradients of the objective function $\Phi$ and one therefore has to differentiate the above matrix operations. In robust settings the objective function often requires higher order derivatives of matrix operations. If there are no constraints in the parameter estimation, the above expression simplifies to
\begin{subeqnarray}
\Phi (C) &=& \trace ( C) = \trace( (J^T J)^{-1}) \label{eqn:unconparam} \;.
\end{subeqnarray}
 The sequence of operations needed in the reverse mode of AD for Eqn. (\ref{eqn:unconparam}) is shown in Table \ref{tab:matrix_ad}. This also motivates the test function in Sect. \ref{sec:performance} which is part of the sequence in Table \ref{tab:matrix_ad}.

\section{Univariate Taylor Propagation on Matrices} \label{sec:utpm}
There are two possibilities how to differentiate matrix operations: Either one regards matrices as two-dimensional arrays and differentiates the linear algebra algorithms, or one considers matrices as elementary objects and applies matrix calculus.
Using UTPS on the first possibility results in the following formal procedure:
\begin{eqnarray}
\left[
\begin{matrix}
[Y_{11}] &\dots& [ Y_{1M_Y}] \\
\vdots & \ddots & \vdots \\
[Y_{N_Y1}] & \dots &[Y_{N_Y M_Y}] \\ 
\end{matrix}
\right] &=&
F
\left(
\left[
\begin{matrix}
[X_{11}] &\dots& [X_{1 M_X}] \\
\vdots & \ddots & \vdots \\
[X_{ N_X 1}] & \dots &[X_{N_X M_X}] \\ 
\end{matrix}
\right] \right)  + \mathcal O(t^{D+1})\;,
\end{eqnarray}
where $N$ is the number of rows and $M$ the number of columns.
A simple reformulation transforms a matrix of truncated Taylor polynomials into a truncated Taylor polynomial of matrices:
\begin{subeqnarray}
\left[
\begin{matrix}
\sum_{d=0}^D X_d^{11} t^d &\dots& \sum_{d=0}^D X_d^{1M} t^d \\
\vdots & \ddots & \vdots \\
\sum_{d=0}^D X_d^{N1} t^d & \dots &\sum_{d=0}^D  X_d^{NM} t^d \\ 
\end{matrix}
\right]
&=&
\sum_{d=0}^D
\left[
\begin{matrix}
X_d^{11} & \dots &X_d^{1M}\\
\vdots & \ddots & \vdots \\
X_d^{N1} & \dots & X_d^{NM}\\
\end{matrix}
\right]
t^D \;.  \label{eqn:utpm}
\end{subeqnarray}
We denote from now on the rhs of Eqn. (\ref{eqn:utpm}) as $[X]$. The formal procudure then reads
\begin{equation}
[Y] = F([X]) + \mathcal O(t^{D+1}) \;,
\end{equation}
which can be treated with matrix calculus. We'd like to call this approach \emph{Univariate Taylor Propagation on Matrices} (UTPM). Notice that even square matrices only form a noncommutative ring.

\section{Reverse Mode on Matrices} \label{sec:utpm_reverse}
Applying the reverse mode to an objective function with matrix argument yields
\begin{subeqnarray}
 \underbrace{\bar \Phi}_{\in \mathbb R} \dd \Phi(\underbrace{Y}_{\in \mathbb R^{N \times M}}) &=& \sum_{n,m} \bar \Phi \frac{ \partial \Phi}{\partial Y_{nm}} \dd Y_{nm} \\
&=& \trace \left( 
\underbrace{ \bar \Phi
\left[
\begin{matrix}
 \frac{ \partial \Phi}{\partial Y_{11}} & \dots & \frac{ \partial \Phi}{\partial Y_{1N}} \\
\vdots & \ddots & \vdots \\
 \frac{ \partial \Phi}{\partial Y_{M1}} & \dots & \frac{ \partial \Phi}{\partial Y_{MN}} \\
\end{matrix}
\right]
}_{=: \bar Y^T \in \mathbb R^{M \times N}}
\underbrace{
\left[
\begin{matrix}
 \dd Y_{11} & \dots &  \dd Y_{1M} \\
\vdots & \ddots & \vdots \\
 \dd Y_{N1}& \dots & \dd Y_{NM} \\
\end{matrix}
\right]
}_{=: \dd Y \in \mathbb R^{N \times M}}
\right) \\
&=& \trace ( \bar Y^T \dd Y ) \;.
\end{subeqnarray}
From that point, one has to successively go backward and find the dependency w.r.t. the arguments $X$ of $Y \equiv Y(X)$.
The reverse mode for the inverse of a matrix $Y=X^{-1}$, transpose of a matrix $Y = X^T$, trace of a matrix $y = \trace(X)$ and the matrix matrix multiplication $Z = X Y$ are given by \cite{giles:08}: 
\begin{align} 
Y = X^{-1}     &:& \trace( \bar Y^T \dd Y )    =& \trace ( \underbrace{- Y \bar Y^T Y}_{=: \bar X^T} \dd X ) \label{eqn:matrix_reverse_inverse} \\
Y = X^T        &:& \trace ( \bar Y^T \dd Y )   =& \trace ( \underbrace{Y}_{=: \bar X^T} \dd X )  \\
y = \trace (X) &:& \bar y \dd \trace(X)        =& \trace( \bar y \Id \dd X ) \\
Z = X Y        &:& \trace( \bar Z^T \dd Z)       =&   \trace \left( Y \bar Z^T \dd X + \bar Z^T X \dd Y \right) \label{eqn:matrix_reverse_multiplication} \;.
\end{align}

\begin{table}[!ht]
\centering
\begin{tabular}{|l c l c l c l|}
\hline
$v_0$ &=& $J$ \phantom{{\Large I}}&&&& \\
\hline
$v_1$ &=& $v_0^T$  &&&&\\
$v_2$ &=&  $  v_1 \cdot v_2 $ &&&&\\
$v_3$ &=& $ (v_2)^{-1} $&&&& \\
$v_4$ &=& $ \trace(v_3) $ &&&& \\
\phantom{lala} &&&&&&\\
\hline
\end{tabular}
\begin{tabular}{|l c l c l c l|}
\hline
$ \bar v_{4}$  &=& $  \bar \Phi $ \phantom{{\Large I}}&&&&\\
\hline
$ \bar v_{3}$  &+=& $  \bar v_{4} \Id$ &&&&\\
$ \bar v_{2}$  &+=& $   - v_{3}^T \bar v_{3} v_{3}^T $ &&&& \\
$ \bar v_{1}$  &+=& $  \bar v_{2} v_{0}^T $ &&&& \\
$ \bar v_{0}$  &+=& $  \bar v_{1}^T v_{2} $ &&&& \\
$ \bar v_{0}$  &+=& $  \bar v_{1}^T $ &&&& \\
\hline
\end{tabular}
\caption{\label{tab:matrix_ad} This table shows how the gradient of Eqn. (\ref{eqn:unconparam}) is computed in the reverse mode of AD. The left side is the function evaluation. All temporary results $v_l$ are saved in memory. They are required in the \emph{reverse sweep} that is shown on the right side. The operations needed in the reverse sweep are defined in Eqn. (\ref{eqn:matrix_reverse_inverse}-\ref{eqn:matrix_reverse_multiplication}). The final derivative can be read from $\bar v_0 \equiv \nabla \Phi$.  }
\end{table}

\section{Higher Order Matrix Derivatives} \label{sec:utpm_higher}
To compute higher order derivatives $\dd^D \Phi$ one can apply UTPM and then use the reverse mode as shown in the previous section. In formulas
\begin{equation}
[ \bar \Phi^T ] \dd \Phi ( [X] ) = \trace ( [\bar X^T ] \dd [ X] ) \;,
\end{equation}
where we have defined $ [ \bar X ] $ and $[ \dd X]$ as Taylor polynomials of matrices as introduced in Sect. \ref{sec:utpm}.

\paragraph{Example: Forward UTPM for the Matrix Inversion}
We want to compute $[X]^{-1}$, where the constant term $X_0 \in \mathbb R^{N \times N}$ is regular. I.e. we have to find $[Y] = [X]^{-1}$ s.t.
\begin{eqnarray*}
1& \stackrel{!}{=} & [X] [Y] =  \left( \sum_{d=0}^D X_d t^d \right) \left( \sum_{e=0}^D Y_e t^e \right)= \sum_{d=0}^D \left(  \sum_{e=0}^d X_e Y_{d-e} \right) t^d + \mathcal O ( t^{D+1}) \;.
\end{eqnarray*}
The Taylor coefficients can now be computed recursively:
\begin{align*}
0:&& X_0 Y_0 &\stackrel{!}{=} 1 && \Leftrightarrow Y_0 = X_0^{-1}\\
1:&& X_0 Y_1 + X_1 Y_0 &\stackrel{!}{=} 0 \\
&& \Leftrightarrow Y_1 &= - X_0^{-1} X_1 Y_0 \\
2:&& X_0 Y_2 + X_1 Y_1 +  X_2 Y_0 &\stackrel{!}{=} 0 \\
&& \Leftrightarrow Y_2 &= - X_0^{-1} \left( X_1 Y_1 + X_2 Y_0 \right) \\
d:&& \sum_{e=0}^d X_e Y_{d-e} &\stackrel{!}{=} 0 && \Leftrightarrow Y_d = - X_0^{-1} \left( \sum_{e=1}^d X_e Y_{d-e} \right) \\
\end{align*}
One can see that the inversion has only to be performed once. If $D$ was large, techniques as used in the fast Fourier transform could be applied. However, typically $D \leq 4$.

\section{Algorithmic Complexity of UTPS vs UTPM} \label{sec:complexity}
In the literature polynomial matrix computations have been thoroughly treated, c.f. e.g. \cite{giorgi:03,cantor:91} and references therein. These publications put more focus on the algebraic complexity theory, are only suitable for large degree $D$ or use an unsuitable complexity measure for our purposes. Here we keep things simple to highlight the difference to the traditional approach in AD theory.

In the theory of AD one traditionally differentiates the algorithms of matrix operations to compute derivatives. For naive implementations of the matrix addition and multiplication the approach of UTPS is equivalent to UTPM when complexity measures neglecting memory movements are used.
More sophisticated algorithms, e.g. the matrix inversion, result in algorithms that are significantly different in the complexity. The computational cost $\OPS$ to compute the whole Taylor series of the matrix inversion is
\begin{eqnarray} 
\OPS ( [X]^{-1} ) &=& \OPS(-X^{-1}) + \sum_{d=1}^D \left( (d+1) \OPS(AB) + (d-1) \OPS(A+B)  \right) \nonumber \\
&=& \OPS(-X^{-1}) + \frac{(D+3)D}{2} \OPS(AB) +\frac{(D-1)D}{2} \OPS(A+B) \;. \nonumber \\ \label{eqn:complexity_utpm}
\end{eqnarray}
The matrix addition is $\mathcal O(N^2)$ and the matrix multiplication is $\mathcal O(N^3)$. We therefore have a computational cost that scales as $\mathcal O( D^2 N^3)$.

Differentiating an algorithm that inverts a matrix requires overloading of the scalar multiplication and addition. The multiplication of two Taylor polynomials needs
\begin{eqnarray*} \label{eqn:complexity_utps}
\sum_{d=0}^D \OPS( \sum_{e=0}^d x_e y_{d-e} ) &=& \sum_{d=0}^D d \OPS(x+y) + (d+1) \OPS(xy) \nonumber \\
&=& \frac{(D+1)D}{2} \OPS(x+y) + \frac{(D+2)(D+1)}{2} \OPS(xy)
\end{eqnarray*}
operations and the addition $ \sum_{d=0}^D \OPS(x_d + y_d) = (D+1) \OPS(x+y) $.  I.e., UTPS needs
\begin{eqnarray}
\OPS ( [X^{-1}] ) &=& \OPS(*,X^{-1}) \left( \frac{(D+1)D}{2} \OPS(x+y) + \frac{(D+2)(D+1)}{2} \OPS(xy) \right) \nonumber  \\
&& + \OPS(+,X^{-1}) \left( (D+1) \OPS(x+y) \right)
\end{eqnarray}
operations in total. The quantities $\OPS(*,X^{-1})$ and  $\OPS(+,X^{-1})$ are the number of multiplications resp. additions in the matrix inversion.
This total operations count can be but does not necessarily has to be the same as Eqn. (\ref{eqn:complexity_utpm}). In the leading powers it is also $\mathcal O(N^3 D^2)$. However, there are several reasons why on a computer there will be significant differences:
The complexity model of counting the operations is inadequate for real computers. One has to consider the cache hierarchy and that a memory access has a latency and a bandwidth that falls behind the speed of the CPU. In the reverse mode of UTPS, many operations that could be computed as one instruction (due to the linearity in the linear algebra) are fetched from the memory. E.g.
for the function $f(X) = X^{-1}$ one has $\OPS(f) = \mathcal O(N^3)$ and therefore the memory requirement using UTPS is ${\rm MEM}(\nabla f) = \mathcal O(N^3)$ but only ${\rm MEM}(\nabla f) = \mathcal O(N^2)$ when using UTPM.
 Also, when UTPS is applied to matrix algorithms, assumptions that were made to make those algorithms fast are no longer valid. E.g. the multiplication of two truncated Taylor polynomials is much more expensive than the addition. Also, due to the unknown degree $D$ it is hard to write tuned algorithms as in ATLAS to avoid cache misses. Of particular importance is also the reduced memory requirement in the reverse mode of AD since using UTPM does not require to tape intermediate results that are used in the linear algebra functions.

\section{Experimental Performance Comparison} \label{sec:performance}
To compare the performance of UTPM to state-of-the-art approaches with UTPS we use an easy but sufficiently complex example for the case $D=1$ has been implemented. The code is available at \cite{sourcecode}. The goal is to compute the derivative $\nabla f \in \R^{N \times N}$ of $f: \R^{N \times N} \rightarrow \R$
\begin{eqnarray}
X & \mapsto & f(X)=  \trace  (X^{-1}) \;.
\end{eqnarray}
\begin{figure}[!t]
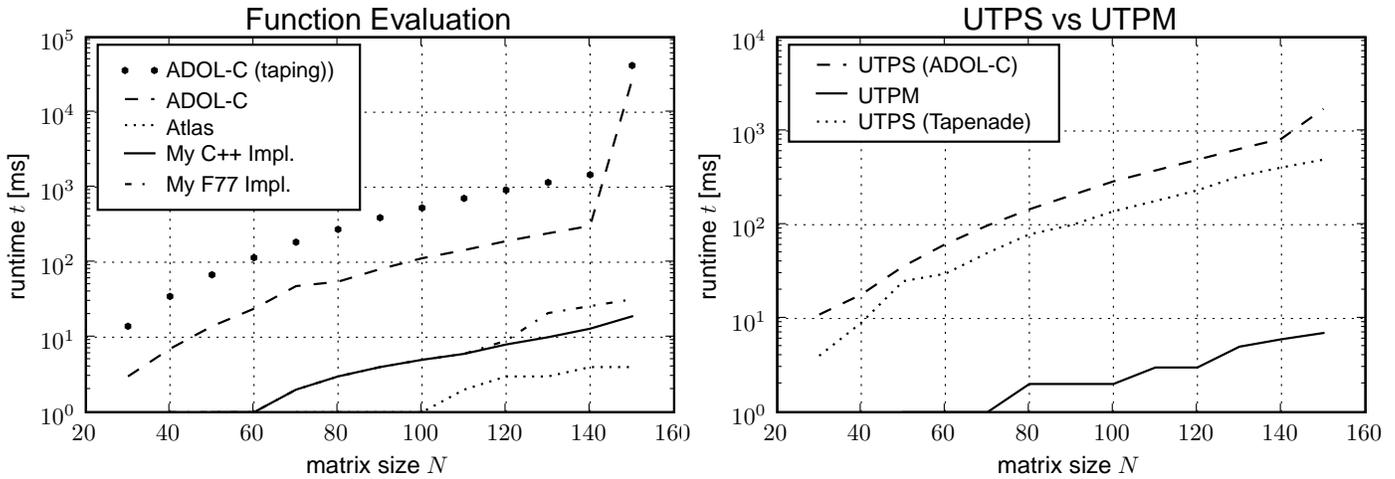

\centering
\includegraphics[width=0.495\textwidth]{pics/function_matrix_ad_vs.epsi}
\includegraphics[width=0.495\textwidth]{pics/gradient_matrix_ad_vs.epsi}
\caption{\label{fig:matrix_ad_vs} This Figure shows a runtime comparison between UTPM implemented using LAPACK and UTPS implemented with ADOL-C resp. UTPS implemented with Tapenade. In the left plot one can see that taping the function $f(X) = \trace (X^{-1})$ with ADOL-C is much slower than a function evaluation. The runtime explosion at $N=150$ is a results from read/write access to the harddisk due to insufficient physical memory. It also shows that our implementation of the QR decomposition is about $5$ to $10$ times slower than LAPACK/ATLAS.
In the right plot one can see that the UTPS approach of both Tapenade and ADOL-C are much slower than UTPM, even if our non-optimal implementation is accounted for.
}
\end{figure}
Since LAPACK code could not readily be differentiated with ADOL-C or Tapenade, we implemented the matrix inversion by QR decomposition using Givens rotations. This code was then taped with ADOL-C and differentiated in reverse mode. \emph{Taping} refers to the process of recording the intermediate values $v_l$ that are needed in the reverse mode of AD. Since Tapenade was not able to differentiate the C++ code necessary for ADOL-C we also implemented it in Fortran 77. The results are depicted and explained in Fig. \ref{fig:matrix_ad_vs}.
%
\footnotesize{ 
\paragraph{Acknowledgments} 
We are grateful to Andreas Griewank for the discussions on computational complexities in AD and to Lutz Lehman for the enlightening discussions and his references \cite{giorgi:03,cantor:91} that are closely related to our approach.

This project is supported by the Bundesministerium f\"ur Bildung und Forschung (BMBF) within the project NOVOEXP (Numerische Optimierungsverfahren f\"ur die Parametersch\"atzung und den Entwurf optimaler Experimente unter Ber\"ucksichtigung von Unsicherheiten f\"ur die Modellvalidierung verfahrenstechnischer Prozesse der Chemie und Biotechnologie) (03GRPAL3), Humboldt Universit\"at zu Berlin.
}

%
%
%

%


\end{document}